\newcommand{\BE}{\begin{equation}}
\newcommand{\EE}{\end{equation}}
\newcommand{\be}{\begin{equation}}
\newcommand{\ee}{\end{equation}}
\newcommand{\bea}{\begin{eqnarray}}
\newcommand{\eea}{\end{eqnarray}}
\newcommand{\atwo}{{a\over 2}}

\documentstyle[twoside,fleqn,espcrc2]{article}

\title{Smooth interpolation of lattice gauge fields by
signal processing methods}

\author{James E. Hetrick
\address{Department of Physics, University of
Arizona, Tucson, AZ 85721, USA}}

\begin{document}

\begin{abstract}
We digitally filter the Fourier modes of the link angles of an abelian
lattice gauge field which produces the Fourier modes of a continuum
$A_\mu(x)$ that exactly reproduces the lattice links through their
definition as phases of finite parallel transport. The constructed
interpolation is smooth ($C^\infty$), free from transition functions,
and gauge equivariant. After discussing some properties of this
interpolation, we discuss the non-abelian generalization of the
method, arriving for SU(2), at a Cayley parametrization of the links
in terms of the Fourier modes of $A^c_\mu(x)$. We then discuss the use
of a maximum entropy type method to address gauge invariance in the
non-abelian case.

AZPH-TH/95-23, hep-lat/9509094
\end{abstract}

\maketitle

\section{Lattices and signal filters}

The explicit relationship between an array of links $U_\mu$ and the
continuum field $A_\mu(x)$ to which it corresponds is an important
ingredient in comparing lattice results with perturbation
theory. While the mapping of continuum interpolations to lattice
configurations is (very) many-to-one, there are various simple algorithms
which give unique continuum fields that serve our purposes to lesser
or greater degree. Probably the simplest of these is the piecewise
constant interpolation where links are $U_\mu = e^{i a A_\mu}$, and
$A_\mu(x)$ is constant along links. Despite its simplistic properties,
this interpolation is a mainstay of lattice gauge theory.

The purpose of this contribution is to explore better interpolations
of a lattice configuration using some ideas borrowed from signal
processing. The starting point is that the interpolating field be {\em
smooth} ($C^\infty$), and that it reproduces the lattice links {\em
exactly} through the definition of parallel transport between $x$ and
$x+a$
\be
U_\mu(x,x+a) = {\cal P} e^{i\int_x^{x+a}{\kern-0.5em dt}
{}~A_\mu(x^1,x^2,...,t,...,x^D)}
\ee
In light of developments in improved actions over the past few
years,
we know that a better representation of the underlying
continuum field is extremely powerful. Knowing $A_\mu(x)$ allows us to
compute (classically) ``perfect'' operators directly in the continuum
such as $F_{\mu\nu}^2$, $F_{\mu\nu}\widetilde F^{\mu\nu}$,
$\partial_\mu A_\mu$, and Det[$D{\kern-0.65em /}~(A_\mu)$], providing
a greatly improved action, topological charge density, gauge fixing
functional, and fermion effective action. It also provides a method of
implimenting chiral fermions {\it del Desperados} \cite{Montvay,BW}.

The interpolation of an unknown function from a discrete set of
measurments is the bread and butter of digital signal processing
\cite{Signal} and I find it useful to think in terms of the following
analogy between the lattice and a digital filter: $$
\matrix{
A_\mu(x) \longrightarrow &\framebox[1.5cm]{\rm lattice}&
\longrightarrow U_\mu~~~~~ \cr
{\rm ~input} \longrightarrow &\framebox[1.5cm]{\rm filter}&
\longrightarrow {\rm output.}
}
$$
Thus, a lattice is nothing but a band limited digital filter of
continuum gauge fields. Ususally a signal filter is represented by its
transfer function $\lambda_k$ in $k$-space: $$
\widehat f_{\rm in}(k) \rightarrow
\framebox[2cm]{$\lambda_k \widehat f_{\rm in}(k)$}
\rightarrow \widehat f_{\rm out}(k).
$$
Similarly the effect of a lattice on $A_\mu(x)$ has a simple
representation, as follows.

\section{An abelian extrapolation}
An abelian link $U_\mu(x,x+a) = e^{i\theta}$ is defined by an angle
$-\pi \leq \theta_\mu(x+\atwo) < \pi$, assigned to the point
$(x^1,x^2,...,x^\mu+{a^\mu\over 2},...,x^D)$. The lattice transfer
function $\lambda_k$ then follows imediately. We first Fourier
transform the angle field
\bea
&&\theta_\mu(x+\atwo) = \cr
&=& \sum_{\vec k} \widehat\theta_\mu(k)
e^{ik_1x^1+\cdots+k_\mu(x+\atwo)+\cdots+ik_Dx^D}\cr
&=& \sum_{\vec k} \int_x^{x+a}{\kern-0.5em dt^\mu}\widehat A_\mu(k)
e^{ik_1x^1+\cdots+k_\mu t^\mu+\cdots+ik_Dx^D}\cr
&=& \sum_{\vec k} \lambda(k_\mu) \widehat A_\mu(k)
e^{ik_1x^1+\cdots+k_\mu(x+\atwo)+\cdots+ik_Dx^D}\cr
&~&\cr
&&\lambda(k_\mu) = {2\sin(ak_\mu/2)\over k_\mu}
\eea
This filter is so common in signal analysis that it has a special
name, {\tt sinc}$(k)$; it arises when the data are integrals of the
underlying signal, rather than samples. A familiar example is a CCD
camera where pixels represent the integrated count of photons over
some length of time.

We then have our extrapolation: the Fourier modes of the continuum
field corresponding to the lattice with angles $\theta_\mu$, are
\be
\widehat A_\mu(k) = {\widehat\theta_\mu(k)\over \lambda(k_\mu)}.
\ee

\subsection{Properties}

\bigskip
\leftline{$\bullet$ Gauge equivariance}
\medskip

It is very easy to show that the above contruction is gauge invariant
when the continuum gauge section $G(x)$ is extracted likewise from its
lattice counterpart by sampling (ie. $\lambda_k = 1$). This gives us a
continuum $\omega(x)$ whose exponential $G(x_i) = e^{i\omega(x_i)}$ at
lattice sites $x_i$ is the gauge transformation applied to the lattice
field. Then the following diagram holds:
\be
\matrix{
{\rm Lattice}& ~ & {\rm Continuum}  \cr
U_\mu(x_i+\atwo) & \longrightarrow & A_\mu(x) \cr
G(x_i) & \longrightarrow & e^{i\omega(x)} \longrightarrow~~ \omega(x)\cr
\downarrow & ~   & 
\downarrow \cr
U^\prime_\mu(x_i+\atwo) & \longrightarrow & A^\prime_\mu(x) =A_\mu(x)
- \partial_\mu \omega(x)
}
\label{gaugeinv}
\ee
ie. the extrapolation of the gauge transformed lattice field
$U^\prime_\mu$, is identical to the continuum gauge transformation
$A^\prime_\mu$ (using the extrapolated section $\omega(x)$) of the
original extrapolation $A_\mu(x)$. This is true at arbitrary $x$.

\bigskip
\leftline{$\bullet$ Topological charge}
\medskip

Since the extrapolated gauge field is $C^\infty$ and periodic, on a
manifold without boundaries it {\em must} have topological charge
$Q=0$, a simple fact maintaining Stokes law.
While some might be unhappy that an extrapolation has $Q=0$ when other
lattice definitions give nonzero charge, I view this as a feature and
not a bug. We should remember that nontrivial topological charge stems
from nontrivial transition functions; if our lattices are periodic,
then our gauge fields should have $Q=0$. Any non-trivial
topological charge arises from a misrepresentation of the
geometric relationship between $A_\mu$ and $F_{\mu\nu}$, and means
that we have allowed singular gauge sections.

The topological charge {\em density} on the other hand, will not be
zero on the extrapolation, and should be a perfectly good measure of
the true topological fluctuations and character of the field. In
particular, $F\widetilde F$ computed from the extrapolated $A_\mu(x)$
can be integrated over many regions small compared to the volume of
the lattice, whose average will then give an accurate measure of
$Q/{\rm Vol}$.

\bigskip
\leftline{$\bullet$ Large gauge transformations}
\medskip
In the diagram (\ref{gaugeinv}), extrapolated gauge transformations
$e^{i\omega(x)}$ are restricted to those with zero winding if we use a
strictly periodic Fourier basis. However two configurations $U_\mu$
and $U^\prime_\mu(x)$ which are related by a large gauge transformation
will only differ in their zero momentum modes, and the
extrapolation to $A_\mu(x)$ and $A^\prime_\mu(x)$ will reproduce this
difference correctly, despite the fact that it cannot produce the
continuum gauge transformation between them.

\section{Nonabelian extrapolation}

There are two essential difficulties in applying the idea of the last
section to nonabelian gauge fields.

\subsection{Nonabelian parallel transport}

...is very nonlinear. The relationship between link and gauge field
is given by the path ordered exponential
\bea
&&U_\mu(x,x+a) = {\cal P}
e^{i\int_x^{\raise+0.5em\hbox{$_{x+a}$}}
{\kern-1.6em dt}~{\bf A}_\mu}\cr
&~&~\cr
&=& 1 + i\int_x^{x+a}{\kern -1.7em dt} {\bf A}_\mu(t) \cr
&-& {1\over 2!}\int_x^{x+a}{\kern -1.7em dt_2}
\int_x^{x+a} {\kern -1.7em dt_1}
\Big\{\Theta(t_2-t_1){\bf A}_\mu(t_2){\bf A}_\mu(t_1) \cr
&&~~~~~~~~~~~~~~~~~+~\Theta(t_1-t_2){\bf A}_\mu(t_1){\bf A}_\mu(t_2) \Big\}\cr
&+& \cdots
\label{Ptrans}
\eea
where $\Theta(t)=(x>0$)?~1~:~0 is a step function, and the
permutations over ordering are explicitly displayed. We must evaluate
the Fourier transform of each term, then do the sum. The main part of
the computation is evaluating integrals of the form
\bea
&&\kern-2em\int_x^{x+a} {\kern -1.7em dt_n}\dots
\int_x^{x+a}{\kern -1.7em dt_2}\int_x^{x+a} {\kern -1.7em dt_1}
\Theta(t_n - t_{n-1})\dots\Theta(t_2 - t_1)\cr
&~&\cr
&&\times e^{ik_nt_n + ik_{n-1}t_{n-1}+\cdots+ ik_1t_1}.
\label{Dfull}
\eea

I have evaluated terms up to third order in ${\bf A}_\mu$ using
\be
\Theta(t) = {1\over 2\pi} \Big[\sum_{n\not=0}{1\over in}e^{it} + t +
\pi\Big],
\ee
on a gauge field with no zero momentum mode. There are
interesting cancellations which give hope that the entire sum can be
done analytically. For example in the second order term
\bea
I_2 &=& \int_x^{x+a}{\kern-1em
dt_2}\int_x^{x+a}{\kern-1em dt_1}
\Theta(t_2 - t_1)e^{ik_2t_2 + ik_1t_1}\cr
&=&{1\over 2\pi i}\Big\{\sum_{n\not=0} {1\over n}\lambda_{k_2+n}
\lambda_{k_1-n} \cr
&& ~~~~~~~+~\Big({\partial\over\partial k_2}
- {\partial\over\partial k_1}\Big)\lambda_{k_2}\lambda_{k_1}\cr
&& ~~~~~~~+~\pi i ~\lambda_{k_2}\lambda_{k_1}\Big\}
e^{i(k_2 + k_1)(x+\atwo)},
\label{Dtwo}
\eea
the sum over $n\not=0$ (done via the $\zeta$
function contour) exactly cancels the partial derivative term, leaving
\be
I_2 = {1\over 2}\lambda_{k_2}\lambda_{k_1}e^{i(k_2 + k_1)(x+\atwo)}
\ee
which is nicely factorized in $k_1$ and $k_2$, linear in each
$\lambda_k$, and centered at $x+\atwo$.

The third order term with two nested theta functions
$\Theta(t_3-t_2)\Theta(t_2-t_1)$ is tedious but straightforward.
While many terms cancel, certain derivatives of
$\lambda_k$ remain in an individual integral of type (\ref{Dfull}),
yielding
\bea
I_3 &=&{1\over 4\pi^2}\Big\{
\lambda^\prime_{k_3}\lambda^\prime_{k_2}\lambda_{k_1} -
\lambda_{k_3}\lambda^\prime_{k_2}\lambda^\prime_{k_1}\cr
&&~+~\pi^2\lambda_{k_3}\lambda_{k_2}\lambda_{k_1}\Big\} e^{i(k_3 +
k_2 + k_1)(x+\atwo)}.
\label{Dthree}
\eea
However the derivative terms {\em do} cancel when summed over the 3!
permutations of the ordering of $\widehat {\bf A}_\mu(k_n)$ leaving only the
last term in (\ref{Dthree}); again factorized in $k_n$,
linear in each $\lambda_k$, and centered at $x+\atwo$.


For SU(2) fields, I have summed the series in (\ref{Ptrans}) based on
the above results and using explicit formulae for the permutations of
products of $\sigma_i$; the result is interesting but
incomplete. {\em Assuming} that permutations cancel the derivative terms at
each order, the coefficient of the last term
$\lambda_{k_n}\lambda_{k_{n-1}}\dots\lambda_{k_1}$ is easily obtained
and produces a series
which evaluates to the unitary matrix
\be
U_\mu(x,x+a) =
{\displaystyle 1 + {i\over 2}\sum_k \lambda_k \widehat {\bf A}_\mu(k)
\over
{\displaystyle 1 - {i\over 2}\sum_k \lambda_k \widehat {\bf A}_\mu(k)}}.
\label{Cayley}
\ee
This is a Cayley representation of $U_\mu$, similar to that recently
espoused by Periwal\cite{Periwal}. This shows that
there must be more terms appearing at higher orders, since the
correct result must faithfully represent
\be
U_\mu(a,b) U_\mu(b,c) = U_\mu(a,c),
\ee
which (\ref{Cayley}) unfortunately does not do. Nonetheless, it seems an
interesting result.

\subsection{Gauge invariance and $k_c$}

In the abelian case we can content ourselves with a frequency cutoff
in $\widehat A_\mu(k)$ and $\widehat \omega(k)$, and due to the
linearity of gauge rotations, the entire construction is
equivariant. In the non-abelian case gauge transformations, which
involve products of fields $G^\dagger(x) A_\mu(x) G(x) - \dots$,
mix Fourier modes of $\widehat G$ and $\widehat A_\mu$ beyond the cutoff
momentum $k_c$.

Again let's turn to engineering for a solution. In situations
where one needs to know the power spectrum at a frequency other than at
discrete values of the measured FFT, the method
of {\em Maximum Entropy} says that the FFT is actually equal to a
Laurent series in the complex plane, which can be
approximated by a Pad\'e polynomial
\be
\widehat f(k) = \sum_x f(x) z^x_k = {1\over {\displaystyle  \sum_p
\alpha_p z_k^p}}
\ee
where $z_k = e^{2\pi i a k}$. We can use the Pad\'e approximation for
$\widehat f(k)$ at any $k: |k|\leq k_c$.

Another ingredient, known as {\em aliasing}, is that when doing an
FFT, the ``measured'' values of $\widehat f_k$ include contributions
from each multiple of $k_c$. Thus,
\be
\widehat F(k) = \sum_{n=-\infty}^\infty \widehat f(k + nk_c)
\ee
where $\widehat F(k)$ is the measured FFT, and $\widehat f(k)$
is the actual spectrum. Can we use the aliasing to get
information about the spectrum at all $k$?

Consider the Pad\'e approximation to $\widehat f(k)$
\be
\widehat f(k) = {{\cal A}(k)\over {\cal B}(k)}
\ee
where ${\cal A}$ and ${\cal B}$ are polynomials. Including the effect
of aliasing this becomes
(here we use lattice units $a = 1$, so that $k_c = L$ and
the fourier frequencies $-{L\over 2}\le k \le {L\over 2}-1$ are integer).
\bea
\widehat F(k) &=& \sum_n {{\cal A}(k - n L)\over {\cal B}(k - n L)} \cr
&=& -\sum_{{\cal R}{\rm es}}\pi {\rm cot}(\pi z){{\cal A}(k - zL)
\over {\cal B}(k - zL)}
\eea
where the residues are computed at the zeros of ${\cal B}(k -
zL)$. If we assume a simple form for ${\cal B}$ such
as\footnote{Engineers will recognize this as a $2L$ pole, lowpass
Butterworth filter.}
\be
{\cal B}(k) = L^{2L} - k^{2L}
\ee
then the poles and residues are easy to find, and we have a linear
system of equations for the coefficients of ${\cal A}(k)$. Since
$A_\mu^c(x)$ is real, take ${\cal A}(k)$ real and even for the
Re$\widehat f$, and imaginary and odd for Im$\widehat f$. ${\cal
A}(k)$ is of order $k^{(2L -2)}$ ensuring finite power in $\widehat
f(k)$. Then, for example
\be
{\rm Re}\widehat f(k) = \sum_{p=0}^L \Big\{\sum_{n=0}^L
 {\pi \cot \pi (k/L -
\phi_n)\over \prod_{m\not= n}(\phi_n - \phi_m) } \phi_n^{2p}\Big\}
A_{2p}
\ee
where $\phi_n = e^{i\pi(n+{1\over 2})/L}$, and $A_{2p}$ is the
coefficient of $k^{2p}$ in ${\cal A}(k)$, with a similar formula in $2p-1$ for
Im$\widehat f(k)$.

\section{Outlook}

I have attempted to outline the issues surrounding a
better interpolation of lattice gauge fields.
In the nonabelian case there is a large technical difficulty in
unraveling the nonlinearity of parallel transport, which however bears
some promise of analytic tractability. On the other issue of
gauge equivariance, it would seem we can at best make a good
approximation to the continuous Fourier spectrum of the continuum
field $A_\mu(x)$. Notice though that this problem arises due to fact
that we've used the Fourier basis.

A major development in signal processing are {\em wavelets},
which unlike Fourier modes are localized both in frequency and space.
Perhaps in an appropriate wavelet basis, the convolution problem might
not be so bad. In fact the simple piecewise linear extrapolation is an
expansion in the Haar wavelet, one of the simplest examples of a
wavelet basis. It is gauge invariant precisely because of its
convolution properties. I hope to explore this approach in the future.

\medskip
I wish to thank Doug Toussaint, Philippe de Forcrand,
Yigal Shamir, and Jan Smit for thoughtful discussions on the above
ideas. This work was supported by the US Department of Energy,
contract DE-FG03-95ER-40906

\end{document}